\documentclass{aa}
\usepackage[varg]{txfonts}
\usepackage{graphicx}
\usepackage{amsmath}
\usepackage{amssymb}
\usepackage{multirow}
\usepackage{lscape}
\usepackage{soul}
\usepackage{longtable}
\usepackage{bm}
\usepackage{hyperref}

\newcommand\hmpc{$h^{-1}$Mpc}

\newcommand{\h}{$h^{-1}$}

\newcommand{\Msol}{$\textrm{M}_\odot$}
\newcommand{\sbr}[1]{_{\textrm{#1}}} 
\DeclareUnicodeCharacter{2003}{}
\DeclareUnicodeCharacter{2212}{}

\begin{document}

\title{Measuring $f\sigma_8$ and BAO scale in the Local Universe: a joint real and redshift space analysis from CosmicFlows-4++}

\author{C. Franco\thanks{camilafranco@on.br}\inst{1,2}, 
H. M.~Courtois\inst{2}, 
A. Bernui\inst{1}
}

\institute{
Observatório Nacional, Rua General José Cristino, 77, São Cristóvão, 20921-400, Rio de Janeiro, RJ, Brazil
\and 
Universit\'e Claude Bernard Lyon 1, IUF, IP2I Lyon, 4 rue Enrico Fermi, 69622 Villeurbanne, France
}

\date{Received A\&A xxx , 2026; Accepted date: }

\abstract {

The large-scale clustering of galaxies encodes both geometric and dynamical information about the Universe. 
The Baryon Acoustic Oscillations (BAO) phenomenon 
provides a standard ruler that constrains the cosmic expansion history, while Redshift Space Distortions (RSD) probe the growth of structure through the peculiar velocity field. 
In this work, we present a joint analysis of BAO and growth rate parameter, $f\sigma_{8}$, in the Local Universe out to $z = 0.1$, using the $65,331$ galaxy distances of CosmicFlows-4++ database. A distinctive property of this catalogue is the availability of real space galaxy positions in addition to the redshift space coordinates. 
Fitting an empirical model to the measurements we obtain $r_{\rm{BAO}}^{\rm{real}} = 132\pm 8\,h^{-1}\,{\rm Mpc}$ in real space, and $r_{\rm{BAO}}^{z} = 139 \pm 7\,h^{-1}\,{\rm Mpc}$ in redshift space, at redshift $z = 0.07$. 
Modeling the enhancement of the correlation function within the Kaiser formalism, we derive a constraint on the growth rate parameter $f\sigma_8 = 0.344 \pm 0.105$. 
This analysis demonstrates how the combination of real and redshift space clustering measurements enables a simultaneous probe of important observables of the large-scale structure. Their joint detection in the same dataset, therefore, provides a self-consistent view of the structure and evolution of the Local Universe.
This study may be used for consistency analyses of upcoming surveys, as DESI and 4MOST, that will also provide data in both real and redshift space.
}


\titlerunning{Local Universe BAO and $f\sigma_{8}$}
\authorrunning{Franco et al.} 
\maketitle
\nolinenumbers

\section{Introduction}
The large-scale distribution of galaxies in the Universe encodes a wealth of information about the cosmic expansion and the dynamical processes governing the growth of structure~\citep{Pezzotta2017,Bautista2018,Haude2019,Marques2020,Aubert2022}. In the middle of this, two important observational probes extracted from galaxy clustering play a relevant role in modern cosmology: the baryon acoustic oscillations (BAO) and the redshift space distortions~\citep[RSD;][]{peebles_large-scale_1980,Cole2005,Eisenstein1998}. Together, these features provide complementary probes of the underlying cosmological model.

The BAO signal originates from acoustic waves propagated in the tightly coupled photon-baryon plasma of the early Universe. 
When 
decoupling occurred, this oscillatory pattern froze in the distribution of matter, leaving a characteristic excess of galaxy pairs separated by a scale corresponding to the comoving sound horizon at the drag epoch. 
This scale acts as a standard ruler in cosmology and is being widely used to constrain the expansion history of the Universe~\citep{Peebles1970, Sunyaev1970, Eisenstein1998, Eisenstein2007, DESI2025}. Measurements of the BAO feature across a wide range of redshifts have contributed to the observational evidence for the current cosmological model, $\Lambda$CDM, as it serves as a robust means to estimate the expansion history of the Universe~\citep{Bassett2010,Cole2005,Eisenstein2005,DiValentino2025,Avila2025,Ribeiro2026,Avila2024,deCarvalho2021,Sousa-Neto2025}.

In addition to this, the large-scale clustering of galaxies also carries dynamical information through the RSD. Once galaxy distances are typically inferred from redshifts, the observed radial positions are perturbed by peculiar velocities induced by gravitational clustering. These velocity fields introduce inhomogeneities in the observed clustering pattern, enhancing the correlation amplitude along the line-of-sight on large scales~\citep{Kaiser1987,Hamilton1998}. In linear theory, this effect depends on the logarithmic growth rate of cosmic structure, as 
\begin{equation}
    f(a) \equiv \frac{d\ln{D}}{d\ln{a}} \,,
\end{equation}
where $D(a)$ is the linear growth factor~\citep{Strauss1995}. Measurements of the quantity $f\sigma_{8}$, which combines the growth rate with the amplitude of matter fluctuations, therefore, provide a probe of the growth of structure and a powerful test of cosmological models and theories of gravity~\citep{Kaiser1987, Hamilton1998, Guzzo2008, Huterer2015,Nunes2021, Bessa2022}.

A challenge in RSD analyses is the degeneracy between the growth rate and the galaxy bias~\citep{Hudson2012}. In most galaxy surveys, the observed clustering amplitude depends on the combination $b\sigma_{8}$, where $b$ is the linear galaxy bias; however, since only redshift space positions are available, RSD measurements typically constrain the parameter
\begin{equation}
    \beta = \frac{f}{b} \,,
\end{equation}
leaving the degeneracy only partially broken~\citep{Kaiser1987, DiPorto2012,DiPorto2012b,Pezzotta2017,Aubert2022,Hawken2012}. Consequently, most analyses rely on additional observables in order to disentangle these quantities~\citep{Song2009,Blake2011,Erdogdu2006,Basilakos2006,Kocevski2006,Avila2021,Avila2022}.

In this work, we explore a complementary approach made possible by the availability of reconstructed real space galaxy positions in the Local Universe provided by the CosmicFlows-4++ catalogue~\citep[hereafter, CF4++;][]{Courtois2023,Tully2023}. The data is composed of both observed redshifts and distance estimates derived from a variety of calibrated distance indicators, combined with large-scale reconstructions of the density and velocity fields. As a result, the catalogue allows galaxy clustering to be studied simultaneously in real and in redshift space using the same sample.

This unique feature offers a direct way to separate the contributions of galaxy bias and gravitational growth. In real space, the two-point correlation function traces the spatial distribution of matter modulated by galaxy bias, providing a constraint on $b\sigma_{8}$. On the other hand, in redshift space, the same clustering signal is modulated by the peculiar velocities, whose effect depends on $f\sigma_{8}$. By comparing the correlation functions of both, it becomes possible to isolate the dynamical contribution of the velocity field and constrain the growth of cosmic structure.

At the same time, the presence of the BAO feature in both real and redshift space provides an opportunity to connect geometric and dynamical probes within a single observational framework. While the BAO scale encodes the geometric imprint of sound waves in the early Universe, RSD probe the late-time dynamics of gravitational growth. Therefore, studying both in a consistent analysis establishes a direct link between the large-scale structure and its dynamical evolution~\citep{Weinberg2013,Aubourg2015}.

In light of the foregoing, in this work we present a joint analysis of BAO and $f\sigma_{8}$ in the Local Universe using the CF4++ galaxy catalogue. We measure the two-point correlation function in both real and redshift space using identical pipeline. The real space clustering is used to determine the galaxy bias amplitude, while the redshift space clustering provides constraints on the growth rate parameter. The BAO feature is detected in both spaces, allowing us to quantify the impact of peculiar velocities on the clustering signal.

This paper is organized as follows: in Section~\ref{sec:datasets}, we describe the observational catalogue and the simulation used in this work. Section~\ref{sec:methodology} presents the methodology adopted to measure the correlation functions and model the BAO and RSD signatures. We present our results in Section~\ref{sec:results}. Finally, our conclusions are summarized in Section~\ref{sec:conclusion}.


	

\section {Observational and simulated datasets}
\label{sec:datasets}
In this section, we outline the observational catalogue that underpins our analysis, the procedures adopted to construct matched random catalogues, 
and the extraction of mock samples from a cosmological simulation. Our objective is to construct a set of observational, random and mock samples that are geometrically and statistically comparable.

\subsection{CF4++ catalogue of galaxies}
\label{sec:cf4}
\begin{figure}[!ht]
    \centering
    \includegraphics[width=0.9\linewidth]{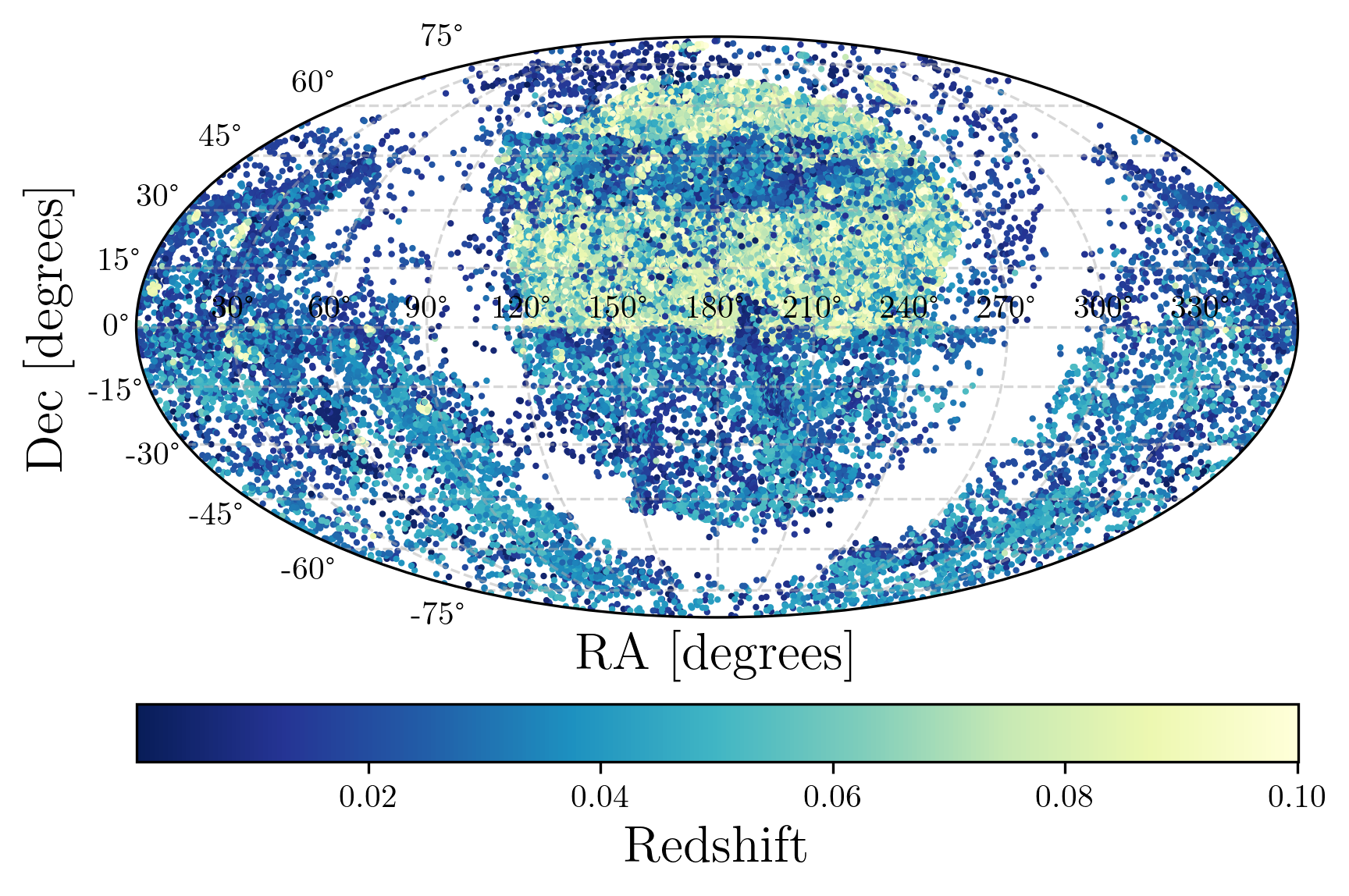}
    \caption{Footprint of the distribution of the CF4++ galaxies in equatorial coordinates. The colorbar represents the redshift range of each galaxy in the catalogue.}
    \label{fig:footprint}
\end{figure}

The primary observational input for this work is the CosmicFlows-4++ catalogue~\citep{Courtois2025}, illustrated in Figure~\ref{fig:footprint}, which provides redshifts and real space distances of $65,331$ galaxies in the Local Universe. These distances estimates are derived from a suite of calibrated distance indicators and assimilated into a large-scale reconstruction of the three-dimension density and velocity fields. The reconstruction methodology is detailed in section 4 of \cite{Courtois2023}; for a more comprehensive explanation on how these calculations are carried out, we refer the reader there.

For the purposes of the present analysis, we employ the catalogue in two representations. In the first, we adopt the reconstruct comoving distances, which, by construction, are free from the line-of-sight displacements induced by peculiar velocities and therefore provide galaxy coordinates in real space. In the second, we convert the observed redshifts into comoving distances assuming a fiducial, spatially flat $\Lambda$CDM cosmology; this representation preserves the characteristic inhomogeneities of redshift space distortions (RSD). 
In both representations, all coordinates are referred to the cosmic microwave background (CMB) rest frame to ensure kinematic consistency between the real and redshift space analyses.

The distinction between these two coordinate systems is central to the interpretation of the clustering signal. In real space, the galaxy distribution traces the underlying matter density field, whereas in redshift space the observed radial coordinate is perturbed by peculiar velocities according to
\begin{equation}
    \label{eq:r2s}
    \vec{s} = \vec{r} + \frac{\vec{v}_{\parallel}}{H_{0}},
\end{equation}
where $\vec{s}$ is the redshift space position along the line-of-sight, $\vec{r}$ is the true comoving distance, $\vec{v}_{\parallel}$ is the peculiar velocity component along the line-of-sight, and $H_{0}$ is the Hubble constant. The comparison between real and redshift space clustering therefore provides a direct probe of the dynamical coupling between the galaxy distribution and the velocity field.

\subsection{Random catalogues}
\label{sec:random}
The random catalogue is construct to match the observational features of CF4++, including its angular footprint, radial selection function, and redshift distribution, but without correlations between the objects. It contains ten times more objects than the data, ensuring that shot noise in $RR(r)$ and $DR(r)$ is subdominant relative to the statistical uncertainties of the data~\citep{Keihanen2019,Avila2024}.

For this work, we implemented the publicly available code \textsc{randomsdss}\footnote{\url{https://github.com/mchalela/RandomSDSS}}\citep{Chalela_RandomSDSS_2021}, which allows us to generate a random redshift distribution from the original measurements while keeping the selection features unchanged. The distances (distance moduli) are assigned to the random points by randomly sampling, from the observed distribution, ensuring that the radial selection function is preserved. The angular distribution is generated following the algorithm from \cite{Franco2024}.

\subsection{MDPL2 mocks}
\label{sec:MDPL2}
We used the MultiDark Planck 2 (MDPL2) simulation, which is part of the MultiDark project, a set of cosmological hydrodynamic simulations~\citep{klypin_multidark_2016}. All simulations in this suite adopt a flat $\Lambda$CDM cosmology with the following parameters: $\Omega_\Lambda=0.692885$, $\Omega_M=0.307115$, $h=0.6777$, $\sigma\sbr{8,linear}=0.8228$, and $n_s=0.96$, in agreement with the \cite{planck_collaboration_planck_2020-values} measurements. MDPL2 follows 3840$^3$ dark matter particles, each with mass $1.51\times 10^9$ \h \Msol. Haloes were identified using Rockstar~\citep{behroozi_rockstar_2012}, which yielding more than 10$^8$ haloes, and halo merger trees were subsequently constructed with ConsistentTrees~\citep{behroozi_gravitationally_2013}. The simulation volume is a periodic cube of side length 1000 $h^{-1}$ Mpc. Our analysis is restricted to the $z=0$ snapshot.
The catalogue was retrieved from the COSMOSIM database\footnote{\url{www.cosmosim.org}}.

We select a region at random from the MDPL2 simulation, ensuring a volume of radius 300 \hmpc, matching the survey depth of CF4++. 
Within this region, we calculate the radial velocity of galaxies, using the Cartesian peculiar velocities provided, in the CMB frame with respect to the origin. 
Additionally, we convert the positions of the galaxies to galactic coordinates, and exclude all objects within Galactic latitudes between $-12^{\circ}$ and $10^{\circ}$ to isolate the Galactic plane region. 

The positions are converted to the same angular coordinate system employed for the data and the identical angular mask is applied. Mock redshift space catalogues are obtained by adding the line-of-sight component of the peculiar velocity to the Hubble flow in the $z$-direction, following Equation~\eqref{eq:r2s}, whereas mock real space catalogues retain the original comoving positions. The final selection of objects in each mock is determined according to the same criteria adopted for the observational sample. 

Since the MDPL2 volume greatly exceeds the effective volume of CF4++, it is possible to extract a large number of statistically independent realizations. In this way, we generated $1000$ mock catalogues that reproduce the survey properties and, therefore, this ensemble provides an estimate of the covariance matrix of the clustering measurements, as will be discussed in Section~\ref{sec:methodology}.

\begin{figure*}[!ht]
    \centering
    \includegraphics[width=0.9\linewidth]{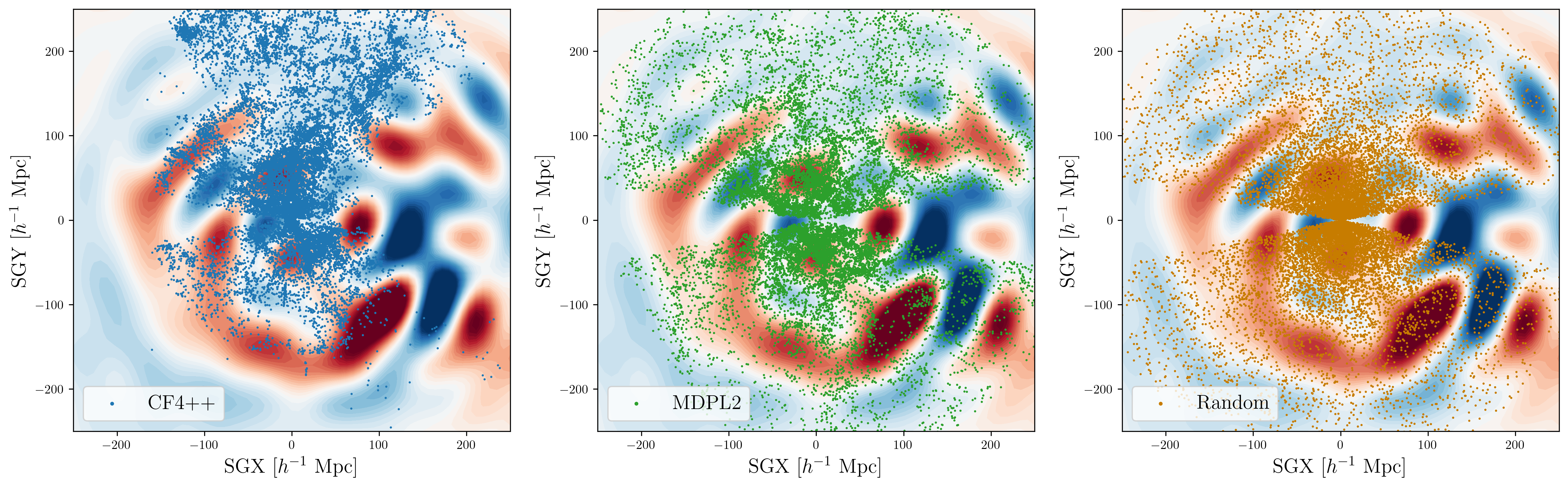}
    \caption{Two-dimensional slice of the large-scale structures in supergalactic coordinates, together with the overdensity field derived from CF4++. Positions from each catalogue are overplotted as the blue, green, and orange points, respectively.}
    \label{fig:slice}
\end{figure*}

\begin{figure*}[!ht]
    \centering
    \includegraphics[width=0.3\linewidth]{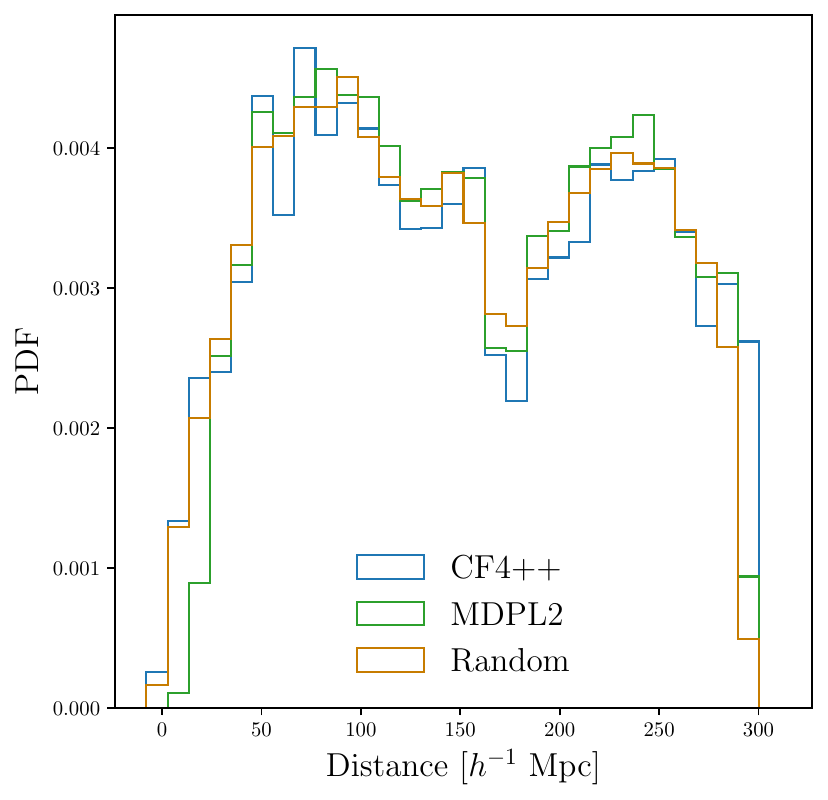}
    \includegraphics[width=0.3\linewidth]{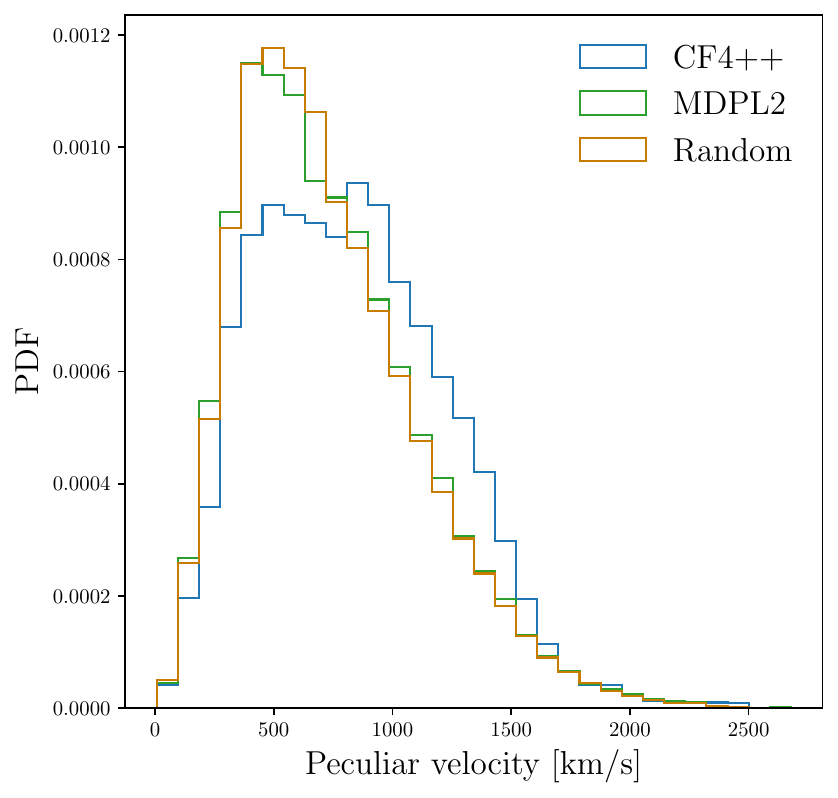}
    \includegraphics[width=0.3\linewidth]{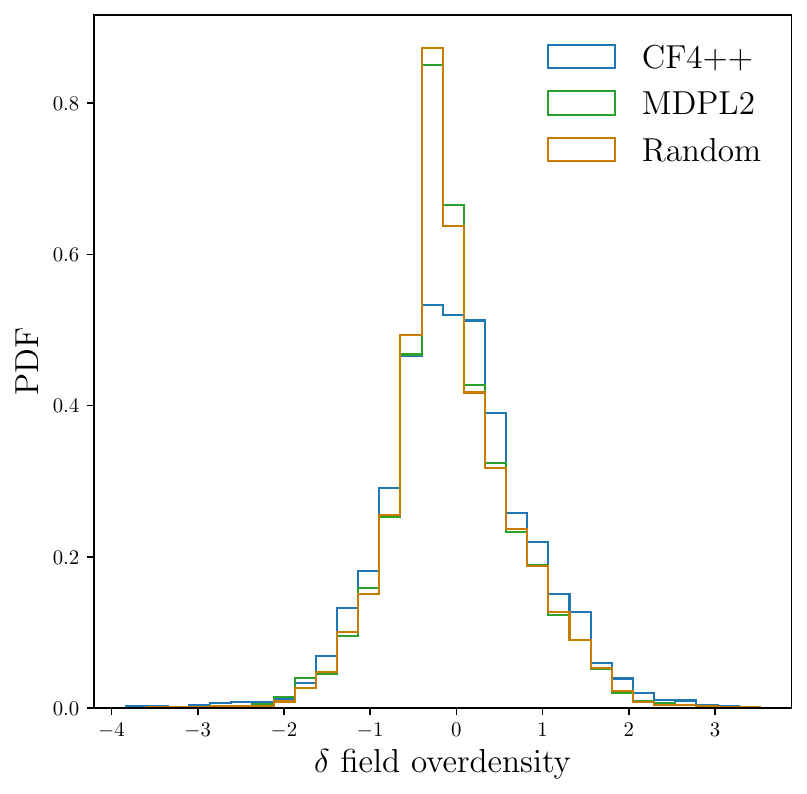}
    \caption{\textbf{Left:} Radial distances of the catalogues. 
    \textbf{Middle:} Peculiar velocity magnitude inferred from the velocity field at the galaxy positions.
    \textbf{Right:} Overdensity field $\delta$ evaluated at the positions of galaxies in each catalogue.
    }
    \label{fig:histograms}
\end{figure*}

Figure~\ref{fig:slice} presents a two-dimensional slice of the large-scale structures in supergalactic coordinates, together with the overdensity field from CF4++. Figure~\ref{fig:histograms} shows the histograms of the radial distances, peculiar velocities, and overdensity matter density field of each catalogue. It is possible to observe the presence of clustered 
structures in both CF4++ and MDPL2 data, while the random distribution does not have any specific signature. 
These features can also be seen when comparing the angular distributions presented in Figure~\ref{fig:ang-footprint}.

\section{Methodology}
\label{sec:methodology}

In this section, we present the methodology adopted to extract the BAO scale and constrain the growth rate of structure from the CF4++ catalogue. Our analysis combines real and redshift space clustering measurements in a framework designed to determine the real-space clustering amplitude, model RSD distortions through the Kaiser formalism to constrain the growth rate of structure, and determine the parameter $f\sigma_{8}$.

\subsection{Measuring the correlation functions}
The spatial distribution of galaxies in the sky is commonly characterized by the two-point correlation function (2PCF), which quantifies the excess probability, relative to a random uniform distribution, of finding a pair of objects separated by a given distance $r$~\citep{Peebles1993}.
To perform our BAO analyses of the CF4++ galaxies, we compute the 2PCF using the Landy-Szalay estimator~\citep[LS;][]{LS1993}, given by
\begin{equation}
    \xi(r) = \frac{DD(r) - 2DR(r) + RR(r)}{RR(r)} \,,
\end{equation}
where $DD(r)$ denotes the number of data-data pairs separated by $r$, $RR(r)$ corresponds to a random-random pairs, and $DR(r)$ represents the cross-correlation between data and random catalogues. This estimator has minimal variance and properly accounts for survey geometry.

All measurements are performed using an identical analysis pipeline for the CF4++, random, and MDPL data. In particular, we apply the same angular mask, radial selection function, binning strategy, and pair-count estimator in all cases, which guarantees that any difference between data and simulations reflects physical effects rather than methodological inconsistencies.

We compute the 2PCF using the public code \textsc{TreeCorr}\footnote{\url{https://rmjarvis.github.io/TreeCorr/_build/html/index.html}}~\citep{Jarvis2015}, adopting $50$ linearly spaced bins in the distance range $[r_{\rm{min}}, r_{\rm{max}}] = [5, 255]\,h^{-1}\,\rm{Mpc}$, i.e, a bin width of $5\,h^{-1}\,\rm{Mpc}$. The binning choice represents a compromise between resolving the BAO feature and maintaining adequate signal-to-noise per bin. We restrict the analysis to separations larger than $5\,h^{-1}\,\rm{Mpc}$ to avoid strongly non-linear scales where the empirical model is not valid. The upper limit of $255\,h^{-1}\,\rm{Mpc}$ is chosen to maximize sensitivity to the BAO peak while minimizing noise from large-scale sample variance.

\subsubsection{Real and redshift space}
In redshift space, galaxy separations are distorted by peculiar velocities along the line-of-sight. We compute the correlation function, $\xi_s(r)$, where the subscript $s$ denotes the redshift space. 
The comoving separation between galaxies $i$ and $j$, with angular separation $\theta_{ij}$, is given by
\begin{equation}
    r_{ij} = \sqrt{\chi(z_i)^{2} + \chi(z_j)^{2} + 2\chi(z_i)\chi(z_j)\cos{\theta_{ij}}} \,,
\end{equation}
where $\chi(z)$ is the comoving distance 
\begin{equation}
    \chi(z) = \frac{c}{H_0}\int_{0}^{z}{\frac{dz^{\prime}}{\sqrt{\Omega_{m0}(1 + z^{\prime})^{3} + (1 - \Omega_{m0})}}} \,,
\end{equation}
assuming a flat $\Lambda$CDM cosmology.

In linear theory, redshift space distortions arise from coherent peculiar velocities generated by gravitational infall, which introduce fluctuations in the correlation function that depend on the growth rate of cosmic structure. 
On the other hand, in real space, peculiar velocities do not affect distances, and the isotropic correlation function, $\xi_r(r)$, directly traces the underlying matter clustering modulated by galaxy bias.
On sufficiently large scales, where density fluctuations remain in the linear regime, this relation provides a direct means to extract the combination $b\sigma_8$~\citep{Franco2025b, Hamilton1998,Desjacques2018}.

As a complementary analysis, we reconstruct the galaxy power spectrum from the measured two-point correlation function via
\begin{equation}
    \label{eq:pk}
    P(k) = 4\pi \int_{0}^{\infty}{r^{2} \xi(r) j_0(kr)dr},
\end{equation}
where $j_0(kr)$ is the spherical Bessel function of order zero. For more details on the power spectrum and the corresponding results, see Appendix\ref{app:pk}.

\subsection{Covariance matrix and uncertainties estimation}
\label{sec:covariance}
Reliable parameter inference requires an accurate description of the statistical uncertainties of the measured correlation function. The uncertainties arise from sampling fluctuations within the observed volume (statistical uncertainties) and cosmic variance associated with the finite realization of the large-scale structure (systematic uncertainties)~\citep{Norberg2009}.

To account for both contributions, we construct a total covariance matrix
\begin{equation}
    C_{\rm tot} = C_{\rm stat} + C_{\rm sys} \,.
\end{equation}
The statistical component, $C_{\rm stat}$, is estimated from the observational sample using jackknife resampling over $N_{jk}$ spatial subvolumes. The resulting covariance estimator is 
\begin{equation}
    C_{ij}^{\rm stat} = \frac{N_{jk} - 1}{N_{jk}} \sum_{k=1}^{N_{jk}}{(\xi_{i}^{(k)} - \bar{\xi}_i)(\xi_{j}^{(k)} - \bar{\xi}_j)} \,.
\end{equation}

The systematic component, $C_{\rm sys}$ is derived from the ensemble of $1000$ MDPL2 mock catalogues described in Section~\ref{sec:MDPL2}. The covariance estimated from the mocks is
\begin{equation}
C_{ij}^{\rm sys} = \frac{1}{N_{\rm mock} - 1} \sum_{k=1}^{N_{\rm mock}}{(\xi_{i}^{(k)} - \bar{\xi}_i)(\xi_{j}^{(k)} - \bar{\xi}_j)} \,.
\end{equation}
The sum of these two matrices provides the estimate of the toal uncertainty affecting the correlation function measurements.

\subsection{Modeling the BAO feature}
To extract the BAO scale, we fit the measured 2PCF using an empirical model widely adopted in the literature~\citep{Sanchez2011, Carnero2012, deCarvalho2018, deCarvalho2021, Avila2024}, 
\begin{equation}
\xi(r) = A + B\,r^{\,\delta} + C\exp{\left[-\,\frac{(r - r_{\rm{BAO}})^{2}}{2\,\Sigma^{2}}\right]} + \frac{D}{r}.
\end{equation}
In this parameterization, $A$ and $B\,r^{\,\delta}$ describe the broadband shape of the correlation function on large scales; $C$ controls the amplitude of the BAO feature; the term $D/r$ captures residual scale-dependent effects at intermediate scales; the Gaussian term centered at the acoustic scale $r_{\rm{BAO}}$ models the BAO peak, where $\Sigma$ accounts for non-linear broadening due to structure formation and peculiar velocities~\citep{Novaes2022,Ross2015,Moon2023,Sanchez2011,Xu2012}.

We constraint the model parameters using a Markov Chain Monte Carlo (MCMC) sampling of the posterior distribution, assuming a Gaussian likelihood of the form
\begin{equation}
    \mathcal{L} \propto \exp{\left(-\frac{1}{2}\chi^{2}\right)},
\end{equation}
where 
\begin{equation}
    \chi^{2} = [\xi_{\rm{model}}(r) - \xi_{\rm{obs}}(r)]^{T} \bm{C}_{\rm tot}^{-1}[\xi_{\rm{model}}(r) - \xi_{\rm{obs}}(r)].
\end{equation}
The priors of the parameters are given in Table~\ref{tab:priors}.
\begin{table}[!ht]
    \centering
    \begin{tabular}{lccc}
        \hline
        \textbf{Parameter} & \textbf{Priors} \\
        \hline
        $A$ & $\mathcal{U}[-5, 5]$\\
        $B$ & $\mathcal{U}[0, 600]$\\
        $C$ & $\mathcal{U}[-1, 1]$\\
        $D$ & $\mathcal{U}[-700, 0]$\\
        $\delta$ & $\mathcal{U}[-5, 5]$\\
        $\Sigma$ & $\mathcal{U}[10, 100]$\\
        $r_{\rm{BAO}}$ & $\mathcal{U}[90, 200]$\\
        \hline
    \end{tabular}  
    \caption{Prior ranges and initial guesses for the power-law parameters.}
    \label{tab:priors}
\end{table}

As a further validation of the analysis pipeline, we compute the correlation function for the random catalogue used in the LS estimator. The result is presented in Appendix~\ref{app:null}.


\subsection{Growth rate estimation}
\label{sec:fs8}


We determine the linear growth rate $f\sigma_{8}$ through a joint analysis of the real and redshift space 2PCF. In this framework, the clustering amplitude, $b\sigma_{8}$, is constrained by the real space correlations, while the dynamical contribution associated with the peculiar velocities is encoded by the redshift space correlations. By modeling both simultaneously, we obtain constraints on the growth rate of cosmic structure, $f\sigma_{8}$~\citep{Song2009}.

From real space, we directly get the 2PCF from galaxy positions. On sufficiently large scales ($r \gtrsim 30\,h^{-1}\,{\rm Mpc}$), one can assume a linear bias as~\citep{Kaiser1987,Desjacques2018}
\begin{equation}
    \label{eq:xi_r}
    \xi_{r}(r) = b^{2} \xi_{m}(r),
\end{equation}
where $b$ denotes the linear galaxy bias and $\xi_{m}(r)$ is the linear matter correlation function, obtained from the Fourier transform of the linear power spectrum, $P_{m}(k)$. The latter is computed using the \textsc{CAMB}~\footnote{\url{https://camb.info/}} code~\citep{camb}, assuming a $\Lambda$CDM model with parameters given by \cite{planck_collaboration_planck_2020-values}.

Peculiar velocities introduce perturbations 
in the observed redshift space 2PCF when galaxy distances are inferred from redshifts. In the linear regime, these anisotropies can be modeled as described by the Kaiser formalism~\citep{Kaiser1987,Hamilton1998}, which predicts the monopole of the redshift space correlation function as
\begin{equation}
    \label{eq:xi0}
    \xi_{0}(r) = \left(b^{2} + \frac{2}{3}bf + \frac{1}{5}f^{2}\right) \xi_{m}(r),
\end{equation}
where $f$ is the growth rate of cosmic structures.

The parameters $b\sigma_{8}$ and $f\sigma_{8}$ are estimated simultaneously through a joint likelihood constructed from the covariance matrices of both 2PCF measurements as
\begin{align}
    \mathcal{L}(b\sigma_8, f\sigma_8) = &-\frac{1}{2}\Big\{[\xi_{r} - \xi_{r,\,{\rm obs}}]^{T}\bm{C}_{\rm tot,\, real}^{-1}[\xi_{r} - \xi_{r,\,{\rm obs}}] \nonumber \\
    & + [\xi_{0} - \xi_{0,\,{\rm obs}}]^{T}\bm{C}_{{\rm tot},\,z}^{-1} [\xi_{0} - \xi_{0,\,{\rm obs}}]\Big\} \,,
\end{align}
where $\bm{C}_{\rm tot,\, real}$ and $\bm{C}_{{\rm tot},\,z}$ denote the total covariance matrix of the real and redshift space correlation functions, constructed as described in Section~\ref{sec:covariance}.

The joint posterior is explored using MCMC, with uniform priors $b\sigma_{8} \in \mathcal{U}[0.1, 2]$ and $f\sigma_{8} \in \mathcal{U}[0.2, 0.6]$. In addition, we restricted the fit to a scale range $r \in [40, 200]\,h^{-1}\,{\rm Mpc}$ to ensure the validity of the linear bias approximation.

\section{Results}
\label{sec:results}
In this section, we present the main results obtained from the CF4++ catalogue. We first compare the real and redshift space correlation functions and quantify the effect of peculiar velocities on the clustering amplitude and BAO scale. We then assess the statistical significance of the acoustic detection and validate the interpretation using MDPL2 mock catalogues processed through the identical pipeline. Furthermore, we compute the growth rate parameter $f\sigma_8$.

\subsection{Correlation functions analyses}
\begin{figure}[!ht]
    \centering
    \includegraphics[width=0.9\linewidth]{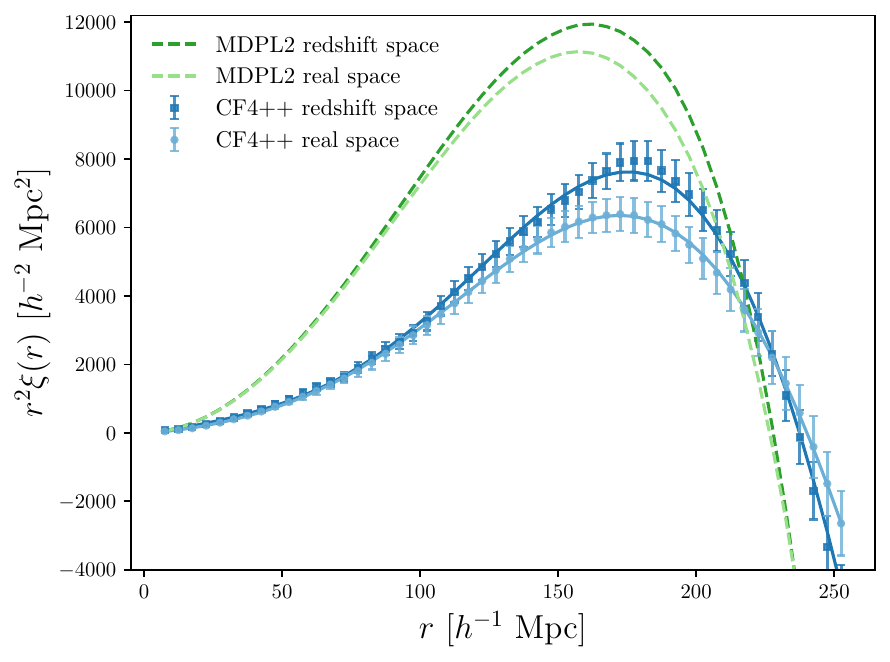}
    \caption{CF4++ two-point correlation functions in real (light blue circles) and redshift (dark blue squares) space, together with the corresponding best-fit model (solid lines). Dashed curves show the mean of $1000$ MDPL2 mocks processed with the same pipeline, also in real (light green) and redshift (dark green) space.}
    \label{fig:corr}
\end{figure}

We first investigate the clustering signal in both real and redshift space. Figure~\ref{fig:corr} shows the measured two-point correlation functions, together with the corresponding best-fitting model described in Section~\ref{sec:methodology}. The comparison highlights the impact of redshift space distortions on both the broadband clustering amplitude and the position of the acoustic peak.

The correlation function is shown in the commonly $r^{2}\xi (r)$ representation, which enhances the visibility of the BAO feature. In this representation, the two cases exhibit a clear systematic difference: in redshift space, the clustering amplitude is enhanced at intermediate and large scales relative to real space. This behaviour is consistent with the Kaiser effect~\citep{Kaiser1987}, in which coherent velocities amplify the clustering along the line-of-sight. Physically, this amplification arises because galaxies falling toward overdense regions acquire peculiar velocities directed to those structures. When distances are inferred from redshifts, these coherent motions compress the apparent distribution of galaxies along the line-of-sight, leading to an increase in the measured clustering amplitude on large scales~\citep{Scoccimarro2004,Hamilton1998}.

Another important point is that the overall shape remains similar in both representations, indicating that peculiar velocities predominantly modulate the amplitude of the clustering signal rather than altering the underlying spatial distribution of matter. The persistence of the acoustic feature in both representations demonstrates that the BAO scale is robust against these velocity distortions.

To quantify the differences, we fit both configurations using the empirical model described in Section~\ref{sec:methodology}\footnote{The MCMC procedure simultaneously constraints the full parameter set of the empirical model. However, since our primary goal is the determination of the BAO scale and derived clustering amplitudes, we focus here on those quantities. The complete posterior summaries for all nuisance and broadband parameters are provided in Appendix~\ref{app:mcmc} for completeness.}. The best-fit curves are overlaid in Figure~\ref{fig:corr}. To assess whether this behaviour is expected within the standard cosmological framework, we measure the same clustering statistics on MDPL2 (represented by the dashed lines in Figure~\ref{fig:corr}). For the real space, we obtain 
\begin{equation*}
    r_{\rm{BAO}}^{\rm{real}} = 132\pm 8\,h^{-1}\,{\rm Mpc}
    \qquad \qquad
    r_{\rm{BAO}}^{\rm{real,\,MDPL2}} = 109\pm 1\,h^{-1}\,{\rm Mpc},
\end{equation*}
while redshift space yields 
\begin{equation*}
    r_{\rm{BAO}}^{z} = 139 \pm 7\,h^{-1}\,{\rm Mpc}
    \qquad \qquad
    r_{\rm{BAO}}^{z\rm{,\,MDPL2}} = 120\pm 0.8\,h^{-1}\,{\rm Mpc}.
\end{equation*}
The corresponding shift, $\Delta r_{\rm{BAO}} = 7 \pm 11\,h^{-1}\,{\rm Mpc}$ in CF4++ data and $\Delta r_{\rm{BAO}} = 11 \pm 1\,h^{-1}\,{\rm Mpc}$ in MDPL2 data, indicates that the peculiar velocities distort the clustering amplitude, but do not alter the physical BAO scale itself. This is import since we aim to constrain the dynamical quantity $f\sigma_{8}$.

We quantify the consistency between data and simulation by computing the tension metric between the estimates,
\begin{equation}
    T_{r_{\rm BAO}} = \frac{|r_{\rm BAO} - r_{\rm BAO}^{\rm MDPL2}|}{\sqrt{\sigma_{r_{\rm BAO}}^{2} + \sigma_{r_{\rm BAO}^{\rm MDPL2}}^{2}}}.
\end{equation}
Real and redshift space scales differ by $2.9\sigma$ and $2.7\sigma$, respectively, from MDPL2, indicating a offset in the absolute BAO scale. In contrast, the BAO shift between real and redshift space is consistent with MDPL2 within $0.4\sigma$, showing that the relative impact of peculiar velocities on the clustering pattern is accurately captured.
Comparing the measurements within the observational data, we find a difference of $0.7\sigma$ between real and redshift space.


Although the absolute amplitude of the correlation function differs between data and mocks, the simulation displays the main qualitative features observed in the data. In particular, the redshift space clustering amplitude exceeds the real space amplitude, reflecting the existence of velocity distortions, as seen in the observational sample. The agreement in the relative behaviour between real and redshift space nevertheless supports the interpretation that the observed differences are primarily driven by physical velocity distortions rather than observational systematics.


\subsection{Growth rate estimation}
\begin{figure}[!ht]
    \centering
    \includegraphics[width=0.9\linewidth]{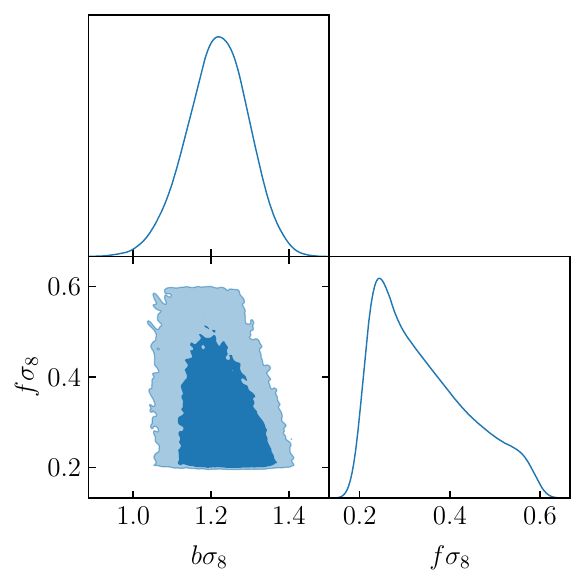}
    \caption{Posterior distributions of the $f\sigma_{8}$ and $b\sigma_{8}$ parameters obtained through the CF4++ data.}
    \label{fig:fs8}
\end{figure}

We constrain the clustering amplitude and the growth rate of cosmic structure through a joint likelihood analysis of the real and redshift space correlation functions, as described in Section~\ref{sec:fs8}. At the effective redshift $z = 0.07$, we obtain
\begin{equation*}
    b\sigma_8 = 1.214 \pm 0.081
    \qquad \qquad
    [b\sigma_8]_{\rm MDPL2} = 0.911 \pm 0.001
\end{equation*}
and a growth rate 
\begin{equation*}
    f\sigma_8 = 0.344 \pm 0.105
    \qquad \qquad
    [f\sigma_8]_{\rm MDPL2} = 0.386 \pm 0.003
\end{equation*}
The fit is performed over the scale range $r \in [20, 200]\,h^{-1}\,{\rm Mpc}$, where the linear bias and Kaiser approximations are expected to provide an adequate description of the clustering signal. The corresponding joint posterior distribution of the CF4++ data is shown in Figure~\ref{fig:fs8}.

As discussed in Section~\ref{sec:fs8}, there is a degeneracy between the parameters $f$ and $b$; and this is evident in the elongated shape of the joint posterior contours of the Figure~\ref{fig:fs8}. 
The inclusion of the real space correlation function helps to break this degeneracy by independently constraining $b\sigma_{8}$. 
However, since the present analysis is restricted to the monopole of the redshift space 2PCF, only part of the anisotropies is captured. 
As a consequence, the sensitivity to $f\sigma_{8}$ remains limited, leading to the relatively large uncertainty in the inferred growth rate.

\begin{figure}[!ht]
    \centering
    \includegraphics[width=0.9\linewidth]{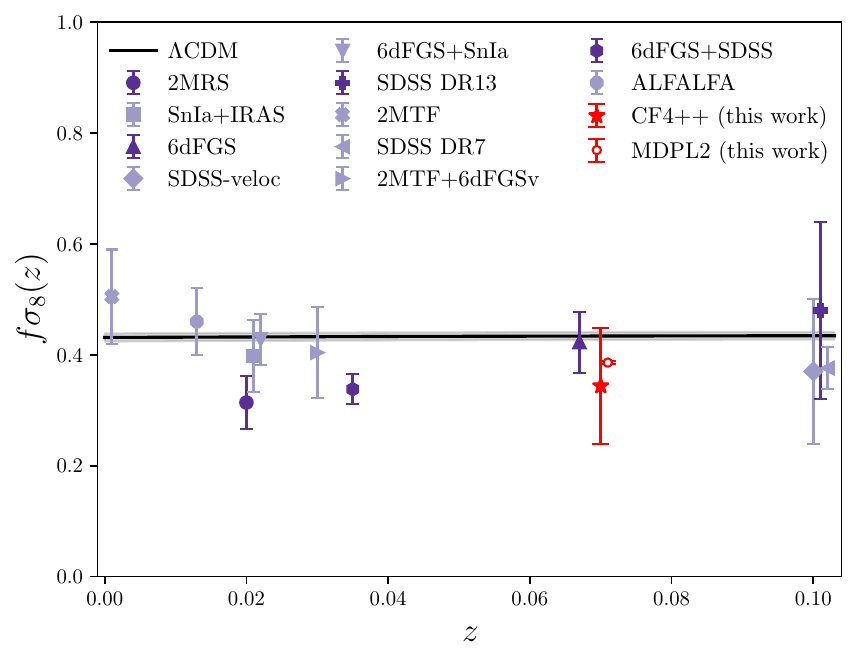}
    \caption{Comparison between measurements of $\sigma_{8}$ as a function of $z$, compared to the prediction of the $\Lambda$CDM model (solid black line). Our measurement using CF4++ is shown as the red star. Measurements within the same redshift have been slightly shifted for better visualization. Points in dark (light) purple are from redshift (real) space datasets.}
    \label{fig:fs8_comparison}
\end{figure}

We compare our result with previous measurements of $f\sigma_{8}$ from different surveys in $z\simeq 0$, summarized in Table~\ref{tab:fs8-values}~\citep{Skara2020,Said2020,Avila2021} and in Figure~\ref{fig:fs8_comparison}. 

\begin{table*}[!ht]
    \centering
    \begin{tabular}{lccc}
        \hline
        \textbf{Dataset} & \textbf{$z$} & \textbf{$f\sigma_8(z)$} & \textbf{Reference}\\
        \hline
        2MRS & $0.02$ & $0.314 \pm 0.048$ & \cite{Davis2011,Hudson2012}\\
        SnIa+IRAS & $0.02$ & $0.398 \pm 0.065$ & \cite{Hudson2012,Turnbull2012}\\
        6dFGS & $0.067$ & $0.423 \pm 0.055$ & \cite{Beutler2012}\\
        SDSS-veloc & $0.1$ & $0.370 \pm 0.130$ & \cite{Feix2015}\\
        6dFGS+SnIa & $0.02$ & $0.428 \pm 0.046$ & \cite{Huterer2017}\\
        SDSS DR13 & $0.1$ & $0.48 \pm 0.16$ & \cite{Feix2017}\\
        2MTF & $0.001$ & $0.505 \pm 0.085$ & \cite{Howlett2017}\\
        SDSS DR7 & $0.1$ & $0.376 \pm 0.038$ & \cite{Shi2018}\\
        2MTF+6dFGSv & $0.03$ & $0.404 \pm 0.0815$ & \cite{Qin2019}\\
        6dFGS+SDSS & $0.035$ & $0.338 \pm 0.027$ & \cite{Said2020}\\
        ALFALFA & $0.013$ & $0.46 \pm 0.06$ & \cite{Avila2021}\\
        \hline
    \end{tabular}  
\caption{Data compilation of $f\sigma_8$ measurements from several surveys and diverse cosmic tracers, with $z \lesssim 0.1$.}
    \label{tab:fs8-values}
\end{table*}

The evolution of the growth rate in the standard model can be approximated, in linear theory, by
\begin{equation}
    f(z) \simeq \Omega_{m}(z)^{\gamma},
\end{equation}
with $\gamma \simeq 0.55$ and the matter density parameter, $\Omega_{m}(z)$~\citep{Linder2007,Linder2020,Basilakos2012, Oliveira2024}. The corresponding prediction for the combination is, then, 
\begin{equation}
    f\sigma_{8} =  \sigma_{8}\,D(z)\,\Omega_{m}(z)^{\gamma},
\end{equation}
where $D(z)$ is the linear growth factor normalized to unity at $z = 0$. Therefore, we can estimate the expected value of the standard model at the same redshift as our CF4++ sample as $[f\sigma_{8}]_{\Lambda {\rm CDM}} = 0.434 \pm 0.007$.

Comparing the inferred $f\sigma_{8}$ values with those obtained from the $\Lambda$CDM model and using the tension metric, we quantify the deviation and find that the observational data is within $0.9\sigma$ from the cosmological model expectation, indicating that our result is consistent with the $\Lambda$CDM framework. For completeness, CF4++ and MDPL2 agree on $0.4\sigma$.

As a quick sanity check, we consider the ratio of the redshift space monopole, Equation~\eqref{eq:xi0}, to the real space, Equation~\eqref{eq:xi_r}, 2PCF. 
Provided the determination of $b\sigma_{8}$, one may define the ratio~\citep{Strauss1995}
\begin{equation}
    \label{eq:ratio_xi}
    R(r) \equiv \frac{\xi_{0}(r)}{\xi_{r}(r)} = 1 + \frac{2}{3}\frac{f}{b} + \frac{1}{5}\left(\frac{f}{b}\right)^{2} \,.
\end{equation}
This ratio encapsulates the first order effects of gravitational infall. By inverting this equation, we obtain a preliminary estimate of $f/b$, which can be multiplied by $b\sigma_{8}$ obtained previously, yielding a quadratic equation for $f/b$,
\begin{equation}
    \frac{1}{5}\left(\frac{f}{b}\right)^{2} + \frac{2}{3}\frac{f}{b} + (1 - R) = 0.
\end{equation}

We measure this ratio over the range $r \in [20, 200]\,h^{-1}\,{\rm Mpc}$, with $b\sigma_8$ drawn from a Gaussian distribution with its measured uncertainty. We retain the positive solution, $f/b = 0.172$, and combine it with $b\sigma_{8} = 1.7 \pm 0.1$, calculated through Equation~\eqref{eq:xi_r}, to obtain 
$f\sigma_{8} = 0.318 \pm 0.212$, 
at redshift $z = 0.07$. 
The uncertainty was estimated through a Monte Carlo procedure, where we recompute $R(r)$ for each realization and then take the standard deviation of the resulting $f\sigma_{8}$ distribution.

This value is consistent with that one obtained from the full joint likelihood analysis within $0.1\sigma$, providing an independent validation of our main result, although this diagnostic is not intended as a precise measurement.





\section{Conclusions}
\label{sec:conclusion}

In this work, we presented a joint analysis of baryon acoustic oscillations and the growth rate of cosmic structure in the Local Universe using the CosmicFlows-4++ catalogue. The key feature of this dataset is the availability of reconstructed real space distances in addition to observed redshift space coordinates, allowing the galaxy clustering signal to be measured in both using the same galaxy sample and analysis pipeline.

We measured the two-point correlation function and detected the BAO feature in both representations of the galaxy distribution. 
Fitting the empirical model described in Section~\ref{sec:methodology} to the measured correlation functions, we obtain a BAO scale of $r_{\rm{BAO}}^{\rm{real}} = 132\pm 8\,h^{-1}\,{\rm Mpc}$ in real space and $r_{\rm{BAO}}^{z} = 139 \pm 7\,h^{-1}\,{\rm Mpc}$ in redshift space. 
The two measurements are consistent within $0.7\sigma$, indicating that the acoustic scale is preserved. 
At the same time, the comparison between both measurements provides a quantification of the impact of peculiar velocities on the observed clustering signal. The detection of the BAO feature in both representations confirms the robustness of the clustering measurement.

The real space clustering provides a direct determination of the clustering amplitude, while the redshift space allows us to isolate the dynamical contribution associated with the peculiar velocity field. From the joint analysis, we obtain a constrain on the growth rate parameter $f\sigma_{8} = 0.344 \pm 0.105$, which is consistent with expectations from the standard cosmological model within the current uncertainties. From MDPL2, we found $[f\sigma_{8}]_{\rm MDPL2} = 0.386 \pm 0.003$.

The key conceptual advantage of this approach lies in breaking the usual degeneracy between galaxy bias and the growth rate. As most large-scale structure analysis are based on redshift space positions, their measurements required marginalization over galaxy bias in order to extract growth rate. The availability of real space galaxy positions in the CF4++ removes this limitation, as it allows the clustering amplitude to be determined independently, providing a route to measuring $f\sigma_{8}$ without bias marginalization.

This methodology is therefore ready to be applied to large datasets in both real and redshift space, such as DESI, 4MOST, and upcoming surveys.




\section*{Data Availability}
The reconstructed density and velocity fields are available as usual at the website of H.~M. Courtois
\url{https://projets.ip2i.in2p3.fr/cosmicflows/}
or upon request if specific help or computational resolution is wanted.

The  data  underlying  this  article are publicly  available  from  the  COSMOSIM  database \url{https://www.cosmosim.org/}, with their respective publications cited in Section \ref{sec:MDPL2}.

\begin{acknowledgements}
CF thanks Coordenação de Aperfeiçoamento de Pessoal de Nível Superior (CAPES) for the financial support. 
HMC acknowledges support from the Institut Universitaire de France and from Centre National d’Etudes Spatiales (CNES), France. 
AB acknowledges a CNPq (Brazil) fellowship. 
The CosmoSim database used in this paper (MDPL2) is a service by the Leibniz-Institute for Astrophysics Potsdam (AIP). The MultiDark database was developed in cooperation with the Spanish MultiDark Consolider Project CSD2009-00064. 
 AI-assisted tools were employed to improve some wording and grammar. 
 This work was carried out using computational resources provided by the Data Processing Center of the National Observatory (CPDON).
 
\end{acknowledgements}

\bibliographystyle{aa} 
\bibliography{bib}

\begin{appendix}

\section{Angular distribution}
\label{app:ang-footprint}
In this appendix, we present the angular footprint of our data, for completeness.

\begin{figure*}[!ht]
    \centering
    \includegraphics[width=0.9\linewidth]{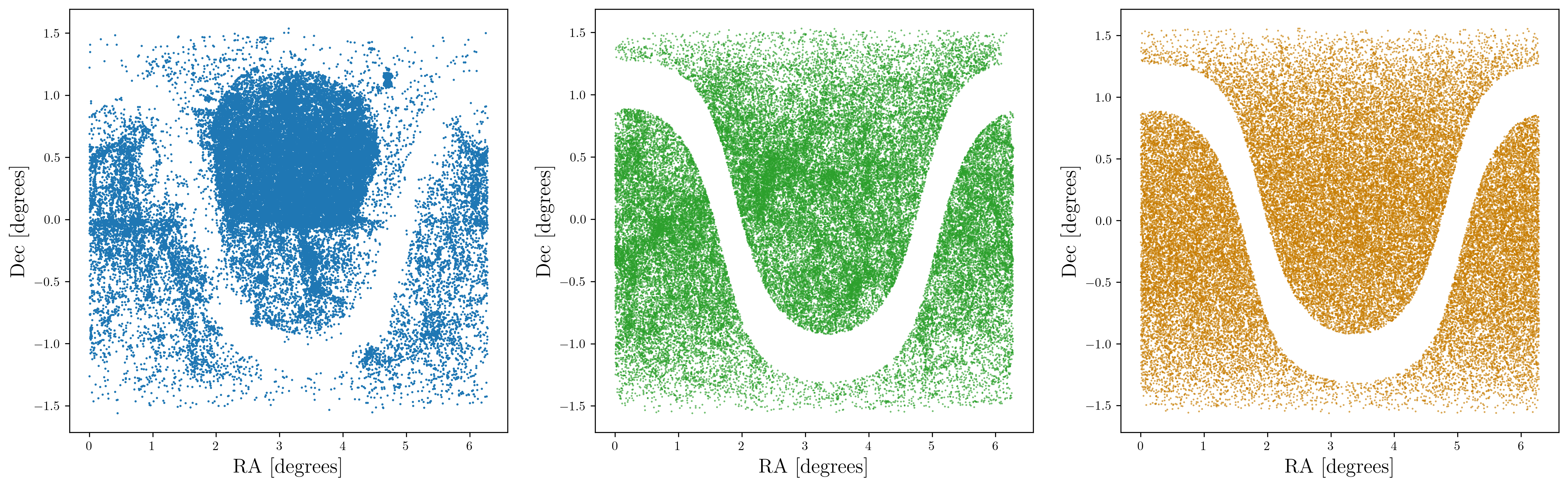}
    \caption{Footprint of the angular distribution of the CF4++ (blue), MDPL2 (green) and random (orange) galaxies in equatorial coordinates.}
    \label{fig:ang-footprint}
\end{figure*}

\section{Matter power spectrum}
\label{app:pk}
The galaxy power spectrum from the measured two-point correlation function is given by Equation~(\ref{eq:pk}). In practice, the integral is evaluated numerically over the finite range spanned by the measured correlation function. Since the observational data provide $\xi(r)$ only within a limited radial interval $[r_{\rm min}, r_{\rm max}]$, we need to restrict the analysis to wavenumbers satisfying $k \gtrsim \pi / r_{\rm max}$ to minimize artifacts induced by the finite integration domain.

To isolate the oscillatory BAO component, we construct a smooth ``no-wiggle'' spectrum, following the analytic fitting formula of \cite{Eisenstein1998}. The smooth transfer function, $T_{nw}(k)$, is computed assuming the fiducial cosmological parameters from \cite{planck_collaboration_planck_2018}. The corresponding smooth power spectrum is given by
\begin{equation}
    P_{nw}(k) \propto k^{n_s}T_{nw}^{2}(k), 
\end{equation}
where $n_s$ is the primordial spectral index.

\begin{figure}[!ht]
    \centering
    \includegraphics[width=0.9\linewidth]{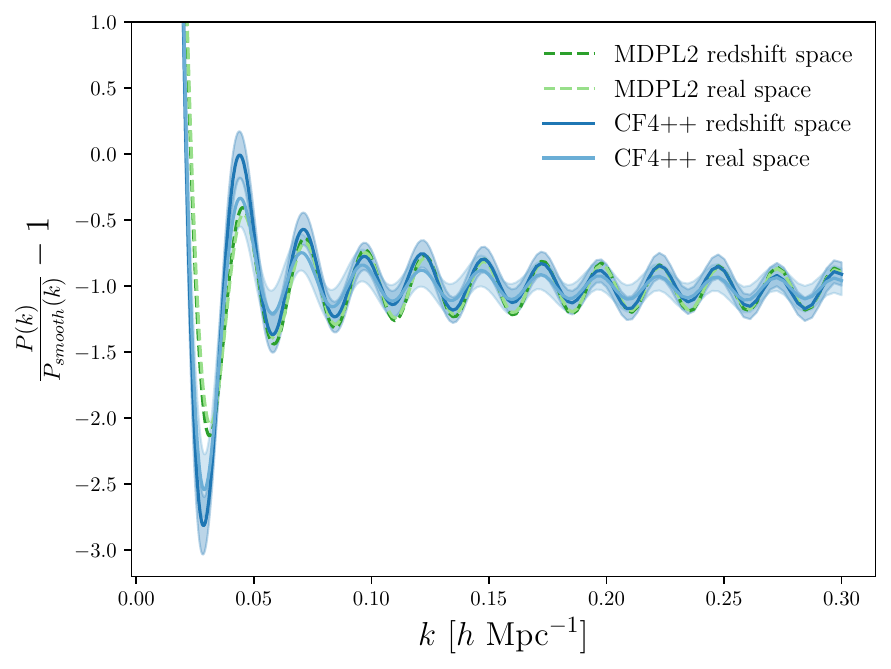}
    \caption{Matter power spectrum in real (green solid line) and redshift (blue solid line) space, together with the corresponding uncertainties range. Dashed curves show the mean of $1000$ MDPL2 mocks processed with the same pipeline.}
    \label{fig:pk}
\end{figure}

The Fourier space representation of the clustering signal provides a complementary view of the BAO phenomenon. Figure~\ref{fig:pk} shows the reconstructed power spectra in both real and redshift space, normalized by a smooth reference spectrum, which isolates the oscillatory BAO component.

While the configuration space correlation function localizes the BAO feature as peak, the power spectrum encodes the same physics as a series of harmonic oscillations whose phase, frequency, and damping are tied to the sound horizon at the baryon drag epoch and the subsequent non-linear evolution of structures. 
In this sense, analyzing the reconstructed power spectrum provides an independent consistency check of the acoustic scale measured in configuration space.

The results show that both the real and redshift space measurements exhibit the expected oscillatory pattern, with peaks occurring at consistent wavenumbers~\citep{Eisenstein2005}. The agreement between the observational and mock measurements further supports that the detected signal corresponds to the BAO feature. In addition, the slight differences between real and redshift space curves reflect the same distortions observed in configuration space.


\section{Null test}
\label{app:null}
\begin{figure}[!ht]
    \centering
    \includegraphics[width=0.9\linewidth]{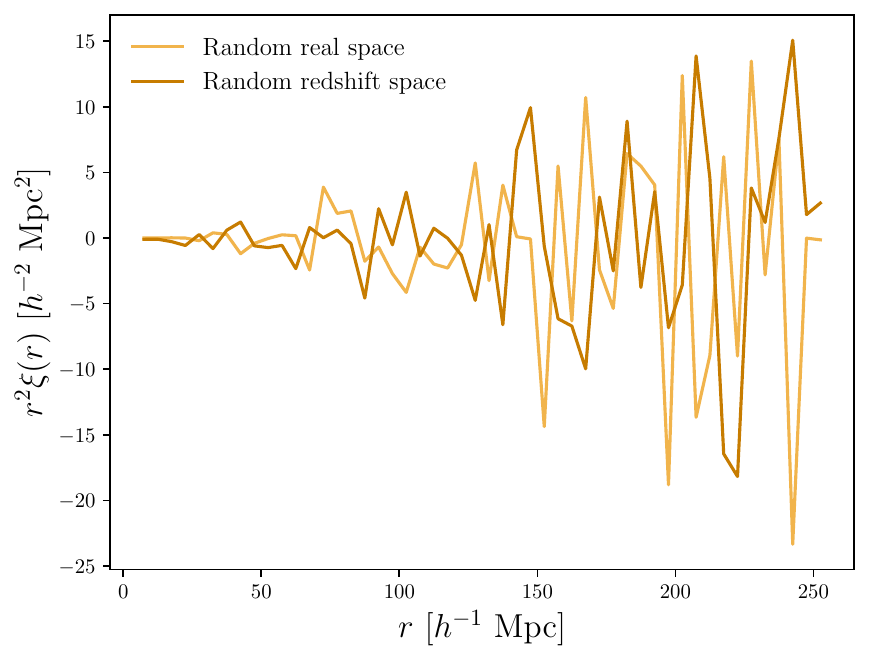}
    \caption{Null test for the random datasets. We perform the 2PCF analysis considering one random as a pseudo-data catalogue. As observed, the curves show no evidence of any signature, just statistical fluctuations consistent with zero.}
    \label{fig:random}
\end{figure}

To check for signatures of artifacts in the random catalogue, we performed a null test to ensure that our results are unbiased. We calculated the 2PCF on a random catalogue as if it were a pseudo-data catalogue in the same way we did with the observational data. The result can be observed in Figure~\ref{fig:random}.

As expected for an unclustered distribution, the correlation function of the random sample is consistent with zero across the full range of separations considered. This confirms that the measured clustering signal arises from the intrinsec spatial distribution of galaxies rather than from artifacts of the survey footprint.

\section{MCMC}
\label{app:mcmc}
\begin{figure*}[!ht]
    \centering
    \includegraphics[width=0.9\linewidth]{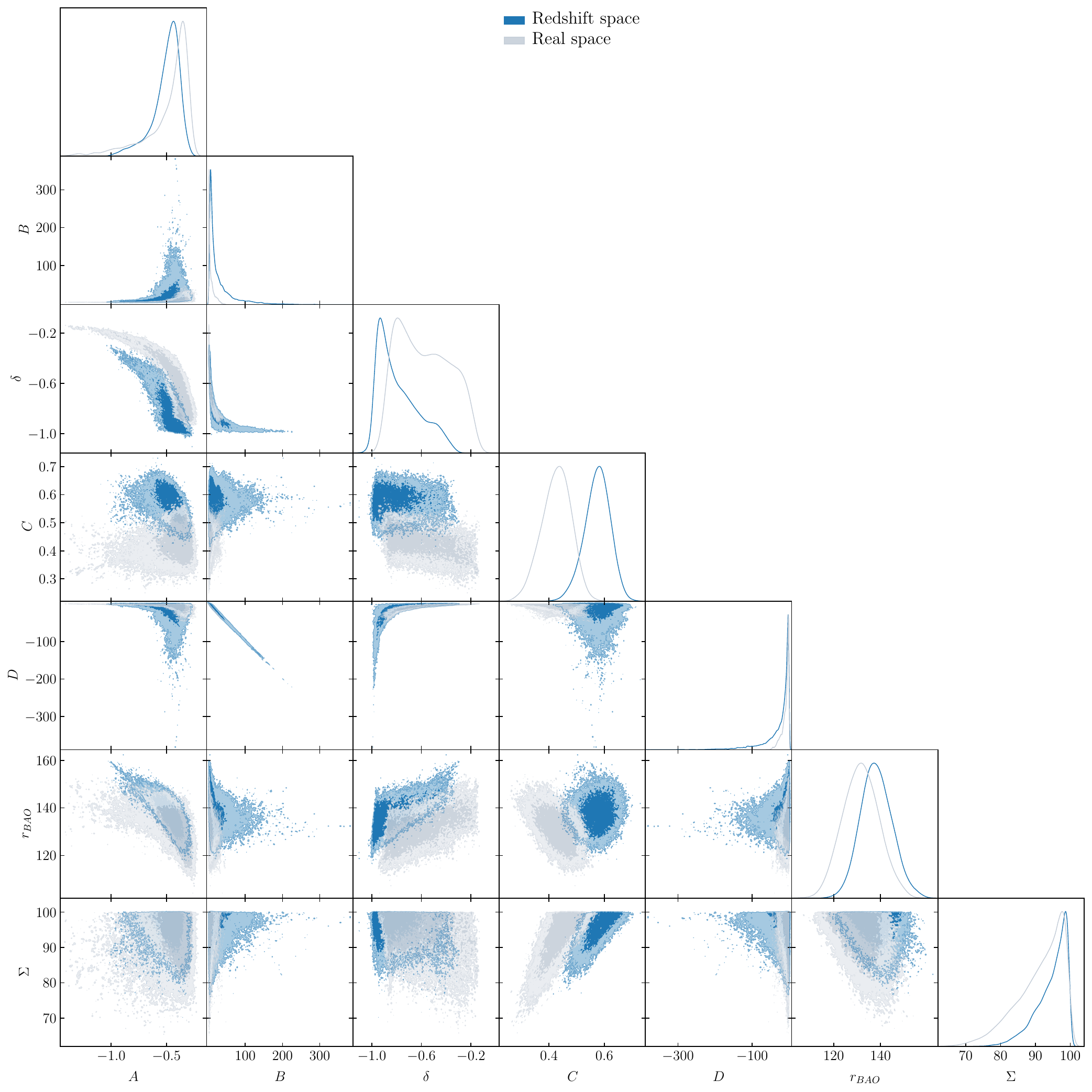}
\caption{Posterior distributions of the BAO model parameters. The blue contours correspond to redshift space, while the gray contours represent real space (see Table~\ref{tab:mcmc}).}
    \label{fig:mcmc}
\end{figure*}

In this appendix, we present the MCMC procedure results, as discussed in Section~\ref{sec:methodology}. Figure~\ref{fig:mcmc} shows the posterior distributions for all parameters, including marginal and joint distributions in both real and redshift space. Summary statistics are listed in Table~\ref{tab:mcmc}

\begin{table}[!ht]
    \centering
    \renewcommand{\arraystretch}{1.5}
    \begin{tabular}{lccc}
        \hline
        \textbf{Parameter} & \textbf{Real space} & \textbf{Redshift space}\\
        \hline
        $A$ & $-0.4093^{+0.0929}_{-0.3291}$ & $-0.4847^{+0.0915}_{-0.1499}$\\
        $B$ & $5.0684^{+13.6970}_{-2.4398}$ & $13.9871^{+29.6859}_{-8.2422}$\\
        $C$ & $0.4277^{+0.0528}_{-0.0591}$ & $0.5781^{+0.0457}_{-0.0524}$ \\
        $D$ & $-5.4995^{+4.1636}_{-15.6996}$ & $-10.8734^{+8.7702}_{-29.9916}$\\
        $\delta$ & $-0.5767^{+0.2906}_{-0.2517}$ & $-0.8017^{+0.2387}_{-0.1322}$ \\
        $\Sigma$ & $93.0924^{+5.1189}_{-9.2664}$ & $95.6834^{+3.1477}_{-5.8138}$ \\
        $r_{\rm{BAO}}$ $[h^{-1}\,{\rm Mpc}]$ & $131.7915^{+8.3863}_{-8.0476}$ & $138.4099^{+7.2792}_{-6.6282}$\\
        \hline
    \end{tabular}  
    \caption{Posterior summaries of the BAO model parameters.}
    \label{tab:mcmc}
\end{table}

From the posteriors, it is possible to observe a shift in several parameters when comparing the two cases, reflecting the impact of RSD on the shape of the correlation function. Parameter $A$, which represents a large-scale offset,
tends to more negative values in redshift space, indicating that RSD modify the overall correlation function at large separations. The broadband component described by the power-law $Br^{\delta}$, shows a degeneracy between $B$ and $\delta$, where the RSD affect mainly the amplitude of this term, as suggested by the distributions. The parameter $C$, which controls the amplitude of the BAO peak modeled by the Gaussian term, is larger in redshift space, consistent with the enhancement of large-scale clustering due to peculiar velocities. At the same time, the width of the BAO peak, $\Sigma$, tends to larger values in redshift space, indicating a smearing of the acoustic feature caused by the velocity distortions. The additional small-scale contributions, given by $D$, presents an asymmetric posterior distribution due to residual variations in the correlation function at small scales. Despite the degeneracies among some parameters, the BAO scale remains well constrained, indicating that the position of the acoustic peak can still be robustly inferred from the data.

\end{appendix}

\end{document}